\documentclass[superscriptaddress, 
reprint,
 amsmath,amssymb,
 aps,
 ]{revtex4-2}

\usepackage{graphicx}
\usepackage{dcolumn}
\usepackage{bm}
\usepackage[colorlinks = true]{hyperref}
\usepackage[range-phrase={\,--\,}, range-units={single}, separate-uncertainty=true]{siunitx}
\usepackage[mathlines]{lineno}
\usepackage{isotope}
\usepackage{stmaryrd}
\usepackage{tikz}
\usetikzlibrary{shapes.callouts}
\tikzset{
    level/.style = {
        ultra thick,
        black,
    },
    connect/.style = {
        dashed,
        black
    },
    label/.style = {
        text width=2cm
    }
}

\begin{document}
\preprint{APS/123-QED}

\title{First direct measurement of the 64.5~keV resonance strength in $^{17}$O(p,$\gamma$)$^{18}$F reaction}

\author{R.\,M.~Gesu\`e}
        \affiliation{Gran Sasso Science Institute, 67100 L'Aquila, Italy}
 \affiliation{INFN, Laboratori Nazionali del Gran Sasso, 67100 Assergi, Italy}
    \author{G.\,F.~Ciani}
    \email{giovanni.ciani@ba.infn.it}
    \affiliation{Università degli Studi di Bari ``A. Moro'', 70125 Bari, Italy}
 \affiliation{INFN, Sezione di Bari, 70125 Bari, Italy}

    \author{D.~Piatti}
    \email{denise.piatti@pd.infn.it}
    \affiliation{Università degli Studi di Padova, 35131 Padova, Italy}
    \affiliation{INFN, Sezione di Padova, 35131 Padova, Italy}

\author{A.~Boeltzig}
    \affiliation{Helmholtz-Zentrum Dresden-Rossendorf, 01328 Dresden, Germany}    

\author{D.~Rapagnani}
 \affiliation{Universit\`a degli Studi di Napoli ``Federico II'', 80125 Naples, Italy}
 \affiliation{INFN, Sezione di Napoli, 80125 Naples, Italy}
 
    \author{M.~Aliotta}
 \affiliation{SUPA, School of Physics and Astronomy, University of Edinburgh, EH9 3FD Edinburgh, United Kingdom}

\author{C.~Ananna}
 \affiliation{Universit\`a degli Studi di Napoli ``Federico II'', 80125 Naples, Italy}
 \affiliation{INFN, Sezione di Napoli, 80125 Naples, Italy}

\author{L.~Barbieri}
\affiliation{SUPA, School of Physics and Astronomy, University of Edinburgh, EH9 3FD Edinburgh, United Kingdom}

\author{F.~Barile}
 \affiliation{INFN, Sezione di Bari, 70125 Bari, Italy}

\author{D.~Bemmerer}
 \affiliation{Helmholtz-Zentrum Dresden-Rossendorf, 01328 Dresden, Germany}

\author{A.~Best}
 \affiliation{Universit\`a degli Studi di Napoli ``Federico II'', 80125 Naples, Italy}
 \affiliation{INFN, Sezione di Napoli, 80125 Naples, Italy}

\author{C.~Broggini}
 \affiliation{INFN, Sezione di Padova, 35131 Padova, Italy}

\author{C.\,G.~Bruno}
 \affiliation{SUPA, School of Physics and Astronomy, University of Edinburgh, EH9 3FD Edinburgh, United Kingdom}

\author{A.~Caciolli}
 \affiliation{Università degli Studi di Padova, 35131 Padova, Italy}
 \affiliation{INFN, Sezione di Padova, 35131 Padova, Italy}

\author{M.~Campostrini}
 \affiliation{Laboratori Nazionali di Legnaro, 35020 Legnaro, Italy}

\author{F.~Casaburo}
\affiliation{Università degli Studi di Genova, 16146 Genova, Italy}
 \affiliation{INFN, Sezione di Genova, 16146 Genova, Italy}

 \author{F.~Cavanna}
 \affiliation{INFN, Sezione di Torino, 10125 Torino, Italy}

\author{P.~Colombetti}
 \affiliation{Universit\`a degli Studi di Torino, 10125 Torino, Italy}
 \affiliation{INFN, Sezione di Torino, 10125 Torino, Italy}

\author{A.~Compagnucci}
 \affiliation{Gran Sasso Science Institute, 67100 L'Aquila, Italy}
 \affiliation{INFN, Laboratori Nazionali del Gran Sasso, 67100 Assergi, Italy}

\author{P.~Corvisiero}
 \affiliation{Università degli Studi di Genova, 16146 Genova, Italy}
 \affiliation{INFN, Sezione di Genova, 16146 Genova, Italy}
\author{L.~Csedreki}
 \affiliation{HUN-REN Institute for Nuclear Research (HUN-REN ATOMKI), PO Box 51, H-4001 Debrecen, Hungary}
 
\author{T.~Davinson}
 \affiliation{SUPA, School of Physics and Astronomy, University of Edinburgh, EH9 3FD Edinburgh, United Kingdom}

 \author{G.\,M.~De Gregorio}
 \affiliation{Università degli Studi di Milano, 20133 Milano, Italy}
 \affiliation{INFN, Sezione di Milano, 20133 Milano, Italy}

 \author{D.~Dell'Aquila}
 \affiliation{Universit\`a degli Studi di Napoli ``Federico II'', 80125 Naples, Italy}
 \affiliation{INFN, Sezione di Napoli, 80125 Naples, Italy}
 
 \author{R.~Depalo}
 \affiliation{Università degli Studi di Milano, 20133 Milano, Italy}
 \affiliation{INFN, Sezione di Milano, 20133 Milano, Italy}
 
  \author{A.~Di~Leva}
 \affiliation{Universit\`a degli Studi di Napoli ``Federico II'', 80125 Naples, Italy}
 \affiliation{INFN, Sezione di Napoli, 80125 Naples, Italy}

\author{Z.~Elekes}
 \affiliation{HUN-REN Institute for Nuclear Research (HUN-REN ATOMKI), PO Box 51, H-4001 Debrecen, Hungary}
 \affiliation{Institute of Physics, Faculty of Science and Technology, University of Debrecen, Egyetem tér 1., H-4032 Debrecen, Hungary}

\author{F.~Ferraro}
\affiliation{INFN, Laboratori Nazionali del Gran Sasso, 67100 Assergi, Italy}

\author{A.~Formicola}
 \affiliation{INFN, Sezione di Roma, 00185 Roma, Italy}

\author{Zs.~Fülöp}
 \affiliation{HUN-REN Institute for Nuclear Research (HUN-REN ATOMKI), PO Box 51, H-4001 Debrecen, Hungary}

\author{G.~Gervino}
 \affiliation{Universit\`a degli Studi di Torino, 10125 Torino, Italy}
 \affiliation{INFN, Sezione di Torino, 10125 Torino, Italy}
\author{A.~Guglielmetti}
 \affiliation{Università degli Studi di Milano, 20133 Milano, Italy}
 \affiliation{INFN, Sezione di Milano, 20133 Milano, Italy}

\author{C.~Gustavino}
 \affiliation{INFN, Sezione di Roma, 00185 Roma, Italy}

\author{Gy.~Gyürky}
 \affiliation{HUN-REN Institute for Nuclear Research (HUN-REN ATOMKI), PO Box 51, H-4001 Debrecen, Hungary}

\author{G.~Imbriani}
 \affiliation{Universit\`a degli Studi di Napoli ``Federico II'', 80125 Naples, Italy}
 \affiliation{INFN, Sezione di Napoli, 80125 Naples, Italy}

\author{M.~Junker}
 \affiliation{INFN, Laboratori Nazionali del Gran Sasso, 67100 Assergi, Italy}

\author{M.~Lugaro}
 \affiliation{Konkoly Observatory, Research Centre for Astronomy and Earth Sciences (CSFK), HUN-REN, and MTA Centre for Excellence, 1121 Budapest, Hungary}
 \affiliation{ELTE E\"otv\"os Lor\'and University, Institute of Physics, 1117 Budapest, Hungary}
 \author{P.~Marigo}
 \affiliation{Università degli Studi di Padova, 35131 Padova, Italy}
 \affiliation{INFN, Sezione di Padova, 35131 Padova, Italy}
 
\author{J.~Marsh}
\affiliation{SUPA, School of Physics and Astronomy, University of Edinburgh, EH9 3FD Edinburgh, United Kingdom}

\author{E.~Masha}
 \affiliation{Helmholtz-Zentrum Dresden-Rossendorf, 01328 Dresden, Germany}
 
\author{R.~Menegazzo}
 \affiliation{INFN, Sezione di Padova, 35131 Padova, Italy}        
        \author{D. Mercogliano}
        \affiliation{Universit\`a degli Studi di Napoli ``Federico II'', 80125 Naples, Italy}
 \affiliation{INFN, Sezione di Napoli, 80125 Naples, Italy}
 \author{V.~Paticchio}
 \affiliation{INFN, Sezione di Bari, 70125 Bari, Italy}

\author{R.~Perrino}
 \altaffiliation[Permanent address: ]{INFN Sezione di Lecce, 73100 Lecce, Italy}
 \affiliation{INFN, Sezione di Bari, 70125 Bari, Italy}

\author{P.~Prati}
 \affiliation{Università degli Studi di Genova, 16146 Genova, Italy}
 \affiliation{INFN, Sezione di Genova, 16146 Genova, Italy}

\author{V.~Rigato}
 \affiliation{Laboratori Nazionali di Legnaro, 35020 Legnaro, Italy}
 \author{D.~Robb}
 \affiliation{SUPA, School of Physics and Astronomy, University of Edinburgh, EH9 3FD Edinburgh, United Kingdom}

\author{L.~Schiavulli}
 \affiliation{Università degli Studi di Bari ``A. Moro'', 70125 Bari, Italy}
 \affiliation{INFN, Sezione di Bari, 70125 Bari, Italy}

\author{R.\,S.~Sidhu}
 \affiliation{SUPA, School of Physics and Astronomy, University of Edinburgh, EH9 3FD Edinburgh, United Kingdom}

\author{J.~Skowronski}
\affiliation{Università degli Studi di Padova, 35131 Padova, Italy}
 \affiliation{INFN, Sezione di Padova, 35131 Padova, Italy}

\author{O.~Straniero}
 \affiliation{INAF-Osservatorio Astronomico d'Abruzzo, 64100, Teramo, Italy}
 \affiliation{INFN, Sezione di Roma, 00185 Roma, Italy}

\author{T.~Szücs}
 \affiliation{HUN-REN Institute for Nuclear Research (HUN-REN ATOMKI), PO Box 51, H-4001 Debrecen, Hungary}

\author{S.~Zavatarelli}
 \affiliation{INFN, Sezione di Genova, 16146 Genova, Italy}
 \affiliation{Università degli Studi di Genova, 16146 Genova, Italy}

\collaboration{LUNA Collaboration}
\noaffiliation

\date{\today}

\begin{abstract}
The CNO cycle is one of the most important nuclear energy sources in stars. At temperatures of hydrostatic H-burning (20 MK $<$ T $<$ 80 MK) the  $^{17}$O(p,$\gamma$)$^{18}$F reaction rate is dominated by the poorly constrained 64.5~keV resonance.
Here we report on the first direct measurements of its resonance strength and of the direct capture contribution at 142 keV, performed with a new high sensitivity setup at LUNA.
The present resonance strength of $\omega\gamma_{(p,
\gamma)}$\textsuperscript{bare} = (30 $\pm$ 
6\textsubscript{stat} $\pm$ 2\textsubscript{syst})~peV is
about a factor of 2 higher than the values in 
literature, leading to a
$\Gamma$\textsubscript{p}\textsuperscript{bare} = (34 $\pm$ 
7\textsubscript{stat} $\pm$ 3\textsubscript{syst})~neV, in 
agreement with LUNA result from 
the (p,$\alpha$) channel. Such agreement strengthen our understanding of the oxygen isotopic ratios measured in red giant stars and in O-rich presolar grains.
\end{abstract}

\maketitle

The CNO cycle releases the energy necessary to sustain the luminosity of red giant, asymptotic giant branch and supergiant stars, and of main-sequence stars with mass $M>1.2$ $M_\odot$. It also powers extended convective zones, such as the convective core of the aforementioned main sequence stars and the convective envelopes of red giants and supergiants. Molecular lines observed in the infrared allows for the measurements of the isotopic composition of selected elements in stars \cite{Hinkel-2016}. The measurements of C, N and O isotopic ratios, in particular, provide a unique tool to understand the interplay between internal nuclear burning and various physical processes producing deep mixing in giant stars. A successful application of this probe in stellar astrophysics requires precise measurements of the burning rates of all the reactions of the CNO cycle. 

\begin{figure}
\centering
\includegraphics[width=\columnwidth]{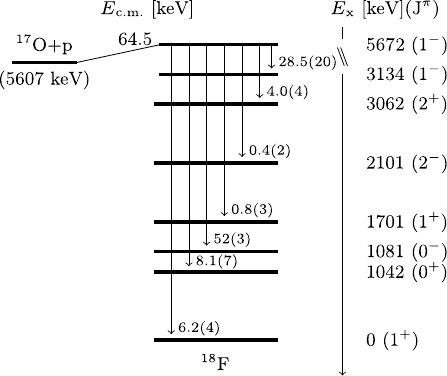}
\caption{Level scheme of \isotope[18]{F} with primary branching ratios as reported in \cite{Tilley_1995}.}
\label{fig:figure1}
\end{figure}

In this paper, we present the first direct measurement of the strength of the narrow resonance at $E$\textsubscript{r}$\,=\,64.5$~keV \footnote{Energies are in the center-of-mass system unless specified differently.} in the $^{17}$O(p,$\gamma$)$^{18}$F reaction ($Q$-value = 5607.1(5) keV \cite{Wang-2021}), corresponding to the $E$\textsubscript{x}$\,=\,5671.6(2)\,$keV \cite{Chafa2007} level in $^{18}$F (\autoref{fig:figure1}). At the temperatures of the stellar hydrostatic H-burning, $T\,\cong\,$20 - 80~MK \cite{Iliadis07-Book}, this nuclear state dominates the rates of the two $^{17}$O destruction channels in the CNO cycle: $^{17}$O(p,$\gamma$)$^{18}$F and $^{17}$O(p,$\alpha$)$^{14}$N. These rates contribute to determine the $^{16}$O/$^{17}$O isotopic ratio observed in giant stars \cite{Harris-1984, Hinkle-2016, Lebeltzer-2019} and in stardust grains that form in the material lost by these stars and are recovered from meteorites \cite{Zinner-2014}.
Attributing the stardust grain origin to a specific scenario is, in fact, particularly sensitive to the choice for the $^{17}$O+p reaction rates \cite{Palmerini2021-Universe}. Reducing their uncertainty would allow us to disentangle the origin of oxide stardust grains from
evolved stars of mass lower than $\sim$1.5 $M_\odot$, thereby proving the existence of extra mixing below the border of the convective
envelope, and/or from evolved stars of mass above $\sim$4 $M_\odot$, where the
CNO cycle occurs at the base of the convective envelope itself \cite{Palmerini2021-Universe, Lugaro-2017NatAs}. 
 
Previously, the \isotope[17]{O}(p,\(\gamma\))\isotope[18]{F} 
reaction was investigated down to $E\,=\,150$~keV through prompt \(\gamma\)-ray detection 
or through offline counting of \isotope[18]{F}
decays from irradiated 
targets (activation method) \cite{Fox2005, Chafa2005, Chafa2007, Hager2012, Kontos-2012, DiLeva-2014, Buckner-2015}. 
Despite experimental efforts, the resonance strength at $E_r = 64.5$~keV in the (p,\(\gamma\)) channel ($\omega\gamma_{(p,\gamma)}$) has never been directly measured.
The presently adopted value was determined through the following relation, after having observed that
for the partial widths 
\(\Gamma_\alpha \gg \Gamma_p , \Gamma_\gamma\) \cite{Blackmon-1995, Buckner-2015, Sergi-2015, Iliadis_2010, Fox2005}:
\begin{equation}
\omega\gamma_{(p,\gamma)} = \frac{(2 J_\mathrm{x} +1)}{(2J_\mathrm{p} + 1) (2J_\mathrm{17O} + 1)} \frac{\Gamma_\mathrm{p} \Gamma_\gamma}{\Gamma_{\alpha}}.
\label{eq:wy}
\end{equation}
The energy $E$\textsubscript{x}$\,=\,5671.6(2)$~keV and spin $J_\mathrm{x}\,=\,1^-$ of the \isotope[18]{F} excited state for the resonance of interest were obtained
by studying the \isotope[14]{N}($\alpha$,$\gamma$)\isotope[18]{F},
\isotope[17]{O}(p,$\gamma$)\isotope[18]{F}, and \isotope[17]{O}(\isotope[3]{He},p$\gamma$)\isotope[18]{F}
reactions \cite{Berka-1977, Bogaert-1989, Chafa2005}.
The $\Gamma_{\alpha} = 130(5)$~eV and $\Gamma_{\gamma} = 0.44(2)$~eV widths were measured via the \isotope[14]{N} + $\alpha$ channels \cite{Mak-1980, Berka-1977}.

The most uncertain quantity in \autoref{eq:wy} is the proton width $\Gamma_\mathrm{p}$, which is estimated from the strength of the $E_r = 64.5$~keV resonance in the (p,$\alpha$) channel,
since \(\omega\gamma_{(p, \alpha)} \propto \Gamma_\mathrm{p}\). 
A discrepancy of a factor 2 and 2.5 exists between the recent result by LUNA \cite{Bruno-2016PhRvL} and values reported by previous direct 
\cite{Blackmon-1995, Buckner-2015} and indirect measurements \cite{Sergi-2015}, respectively, see \autoref{tab:soa}. 
A new independent direct measurement of the \(64.5\)~keV resonance strength in the $^{17}$O(p,$\gamma$)$^{18}$F channel is needed to address this tension. However, due to the low expected count rate, such a direct measurement requires an optimized setup providing both an ultra low background and an extremely high detection efficiency. 

In the following, we present the first direct measurement of the $64.5$ keV resonance strength
performed at the Laboratory for Underground Nuclear Astrophysics (LUNA), 
located at the Laboratori Nazionali del Gran Sasso (LNGS, Italy).

\begin{table*}[]
\centering
\caption{\small{Literature data for the $E$\textsubscript{r} = 64.5~keV resonance in the  $^{17}$O(p,$\gamma$)$^{18}$F reaction: excitation energy $E_{\rm x}$, spin-parity $J^\pi$, partial widths $\Gamma_{\alpha,\gamma,\mathrm{p}}$ and resonance strength $\omega\gamma_{(p,\gamma)}$. For the proton width $\Gamma$\textsubscript{p} and the resonance strength, when explicitly reported, the results corrected for the screening effect are also listed. For the present result the screening correction $f=1.15$ was estimated by applying the adiabatic approximation \cite{Assenbaum-1987} } \label{tab:soa}}
\begin{tabular}{c c c c c c c c c c}
\toprule
    Reference & $E$\textsubscript{x} [keV] & $J^{\pi}$ & $\Gamma_{\alpha}$ [eV] & $\Gamma_{\gamma}$ [eV] & $\Gamma$\textsubscript{p} [neV] & $\Gamma$\textsubscript{p}\textsuperscript{bare} [neV] & $\omega\gamma_{(p,\gamma)}$ [peV] & $\omega\gamma_{(p,\gamma)}$\textsuperscript{bare} [peV] \\
\toprule
\cite{Parker-1968} & & & & 0.45(5) & & & &\\
\cite{Rolfs-1973} & 5669(2) & & & 0.47(10) & & & & \\
\cite{Berka-1977} & & 1$^{-}$ & 200(60) & 0.46(6) & & & \\
\cite{Mak-1980} & & & 130(5) & & & & &\\
\cite{Becker-1982} & & & & 0.45(2) & & & &\\
\cite{Bogaert-1989}  &  5672.57(32) & & & & & & &\\
\cite{Blackmon-1995} & & & & & 22(4) & & &\\
\cite{Hannam-1999} & & & & & 21(2)\footnote{reanalysis of the experimental work by \cite{Blackmon-1995}.} & & &\\
\cite{Chafa2005} & 5671.6(2) & & & & & & &\\
\cite{Buckner-2015, Iliadis_2010, Fox2005} & & & & 0.44(2)\footnote{Average value of the results in \cite{Parker-1968, Rolfs-1973, Berka-1977, Becker-1982}} & &19(3)\footnote{reanalysis of the experimental work by \cite{Blackmon-1995} taking into account the method described in \cite{Hannam-1999} and correcting for the screening effect.} &  & 16(3)  \\
\cite{Sergi-2015} & & & & & & 14(2)\footnote{calculated by present authors starting from the reported $\omega\gamma_{(p,\alpha)}$ and eq.\ref{eq:wy}} &  & 11.8(21) \\
\cite{Bruno-2016PhRvL} & & & & & 40(7) & 35(6) & &\\
Present Work & & & & & 39(9) & 34(8) &  34(8) & 30(6) \\
\toprule
\end{tabular}
\end{table*}

The LUNA deep-underground location guarantees a muon- (and neutron-) background level six (and three) orders of magnitude lower than above ground \cite{Caciolli2009, boeltzig_improved_2018}. 
A detailed description of the setup and the achieved sensitivity can be found in \cite{Skowronski-2023JPhG},
here we only report its main features.

A high-intensity proton beam (average current on target $I\,=\,200$ $\mu$A) 
was provided by the LUNA-400~kV accelerator \cite{Formicola-2003}. The beam was analyzed, collimated and then sent through a copper pipe, extending to a distance of 1~cm from the target.
The pipe was biased to $\mathrm{-}$300 V and kept at liquid nitrogen temperature to work both as secondary electron suppression and as a cold trap. The beam impinged on a water cooled solid target. The target holder and the scattering chamber were made of aluminum to minimize the 
$\gamma$-ray absorption providing an increase of efficiency with respect to previous 
setups of about a factor 4 \cite{Skowronski2023, Skowronski-2023JPhG}. Moreover, the scattering chamber and the target were electrically insulated from the beam line and 
functioned as a Faraday cup for beam current measurement. 

\begin{figure}
\centering
\includegraphics[width=\columnwidth]{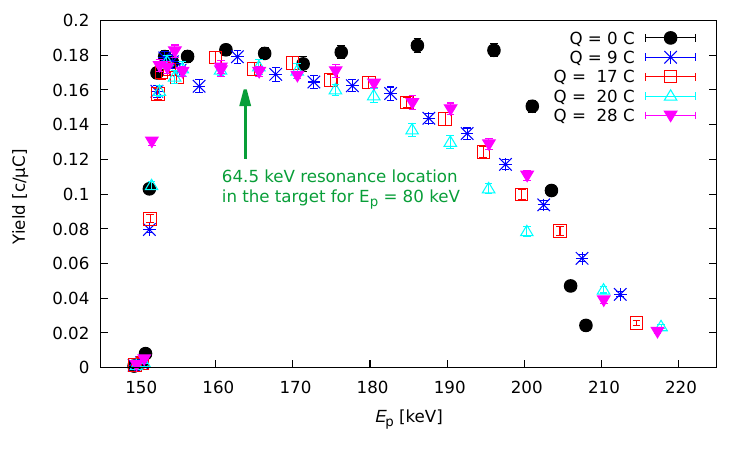}
\caption{Thick-target scans of the $E_r = 143$~keV resonance in the $^{18}$O(p,$\gamma$)$^{19}$F. Scans were acquired after different amounts of
charge deposited on target, and always before and after a long run on the $E_r = 64.5$ keV resonance. Only the statistical uncertainty is plotted. The green arrow shows the beam energy at which the resonance was populated at $E$\textsubscript{p} = 80~keV.}
\label{fig:figure2}
\end{figure}

The targets were produced at LNGS 
by anodic oxidation of tantalum backings, previously cleaned with an acid bath in isotopically $^{17}$O enriched water doped with 4\% of $^{18}$O \cite{Caciolli2012}. 
This procedure was proven to provide targets with a well known stoichiometry, Ta\textsubscript{2}O\textsubscript{5}, and a well defined thickness-voltage relation \cite{Caciolli2012}.
The well known $E$\textsubscript{r} = 143~keV resonance in the $^{18}$O(p,$\gamma$)$^{19}$F channel \cite{best_18opg_2019, Pantaleo-2021PhRvC} was used to characterize the target thickness. Targets with two different thicknesses were used: $\Delta E_{\mathrm{lab}}\,=\,21(1)$ and 53(1)~keV at 143~keV, corresponding to 147 and 378 nm, respectively \cite{SRIM2003}.
To monitor the target degradation during the measurement a resonance scan of the aforementioned resonance was performed periodically, i.e., after every 
$\approx\,10$~C accumulated charge on target. 
The target degradation observed at $E_r = 143$ keV can be directly related to the degradation at
$E_r = 64.5$ keV, since the stopping powers are known and the energy loss at the two energies is nearly the same (within 10\% \cite{SRIM2003}). Targets were replaced after about 25~C of accumulated charge to guarantee the stability of the target stoichiometry within the layer where the 64.5~keV resonance was populated using an $E$\textsubscript{p}$\,=\,80\,$keV beam, see \autoref{fig:figure2}.
 
To characterize and monitor the target isotopic enrichment in $^{17}$O, dedicated runs 
were acquired periodically at $E_\mathrm{p}\,=\,200$ keV, populating the $E$\textsubscript{r} = 183 keV resonance of $^{17}$O(p,$\gamma$)$^{18}$F. 
This resonance has a known strength of \(\omega\gamma_{(p,\gamma)} = ( 1.67 \pm 0.12 )\,\mu\)eV
\cite{DiLeva-2014}. The resulting experimental isotopic abundances of \isotope[17]{O} in the three batches of targets are: 87(1)\%, 72(1)\% and 85(1)\%, where only statistical uncertainties are reported. 

To detect the $\gamma$-rays from the $^{17}$O(p,$\gamma$)$^{18}$F reaction a high efficiency $4\pi$ bismuth germanate oxide (BGO) summing detector was installed around the target and the scattering chamber.
The BGO is segmented into six optically independent crystals,
each read out by a photo-multiplier tube (PMT) and a digital data acquisition chain 
\cite{Skowronski-2023JPhG, boeltzig_improved_2018, boeltzig_23nap_2019, Ferraro-2018PhRvL}.
The energy deposited and the timestamp of each event were recorded and used to 
produce a spectrum of coincident events in different crystals, hereafter referred to as the 
add-back spectrum \cite{boeltzig_improved_2018}. 
The dead time, less than 1\%, was determined using a pulser signal connected to the test input of each preamplifier and to a dedicated acquisition chain. 

Finally, the whole setup was surrounded by a three-layer shielding to further reduce the $\gamma$-ray
background, mainly due to reactions induced by 
environmental neutrons \cite{boeltzig_improved_2018}.
The shield was composed, from inner to outer, of a 1~cm thick borated polyethylene (BPE) layer, a 10~cm thick lead shell and a 5~cm thick BPE cover \cite{Skowronski-2023JPhG}.

A simulation of the setup was developed using the Geant4 framework \cite{Agostinelli_2003}.
The simulation was validated for $\gamma$-ray energies 
up to $7.6$ MeV using
\isotope[137]{Cs} and \isotope[60]{Co} calibrated source and the
$E$\textsubscript{r} = 259~keV 
resonance of the \isotope[14]{N}(p,$\gamma$)\isotope[15]{O} reaction \cite{Adelberger-2011}.
The simulation allowed to characterize the BGO efficiency within 3\% uncertainty \cite{Skowronski-2023JPhG}. 

The data taking covered 4 months, with an overall accumulated charge of 420 C on isotopically enriched $^{17}$O targets and 300 C Ultra Pure Water (UPW) targets to investigate beam induced background, see below for details.
Long runs ($\sim$12~hours each) were performed at $E$\textsubscript{p} = 80~keV to populate the 64.5~keV resonance. Scans of the 143 keV resonance and runs on top of the 183 keV resonance were performed between the long runs to monitor the target degradation.

The 64.5~keV resonance strength is determined from the experimental yield $Y_{\mathrm{exp}}$ using the infinitely-thick-target approximation, i.e., the resonance width (130~eV) is much smaller than the target thickness (53~keV and 21~keV): 
\begin{equation}
Y_{\mathrm{exp}} =\frac{N_{\gamma}}{Q}=\frac{\lambda^2}{2e\epsilon_\mathrm{eff}}\omega\gamma_{(p,\gamma)}\eta W,
\label{eq:yield}
\end{equation}
where \(N_{\gamma}\) is the number of net counts, \(Q\) the 
accumulated charge, 
\(e\) the elementary charge and \(\eta\) the detection
efficiency. The angular distribution term $W$ is 1 since the BGO covers most of the solid angle, $\lambda$ represents the de Broglie wavelength at the center-of-mass resonant energy,
and $\epsilon_\mathrm{eff}$ is the effective stopping power calculated using SRIM-v.13 database \cite{SRIM2003}. 

The yield from the $64.5$ ~keV  resonance was expected to
be less than $0.3$~counts/C \cite{Iliadis_2010}. Therefore defining the region of
interest (ROI) in the gamma energy spectra was critical. The long runs were precisely calibrated up to $E_{\gamma}\,=\,8$~MeV using the 143 keV resonance spectra, acquired before and after long runs allowing to monitor possible gain shift. The regions of interest were determined via a dedicated study of BGO resolution and via simulation of the $E$\textsubscript{x} = 5672~keV de-excitation cascades (\autoref{fig:figure1}).

\begin{figure}
   \centering
   \includegraphics[width=\columnwidth]{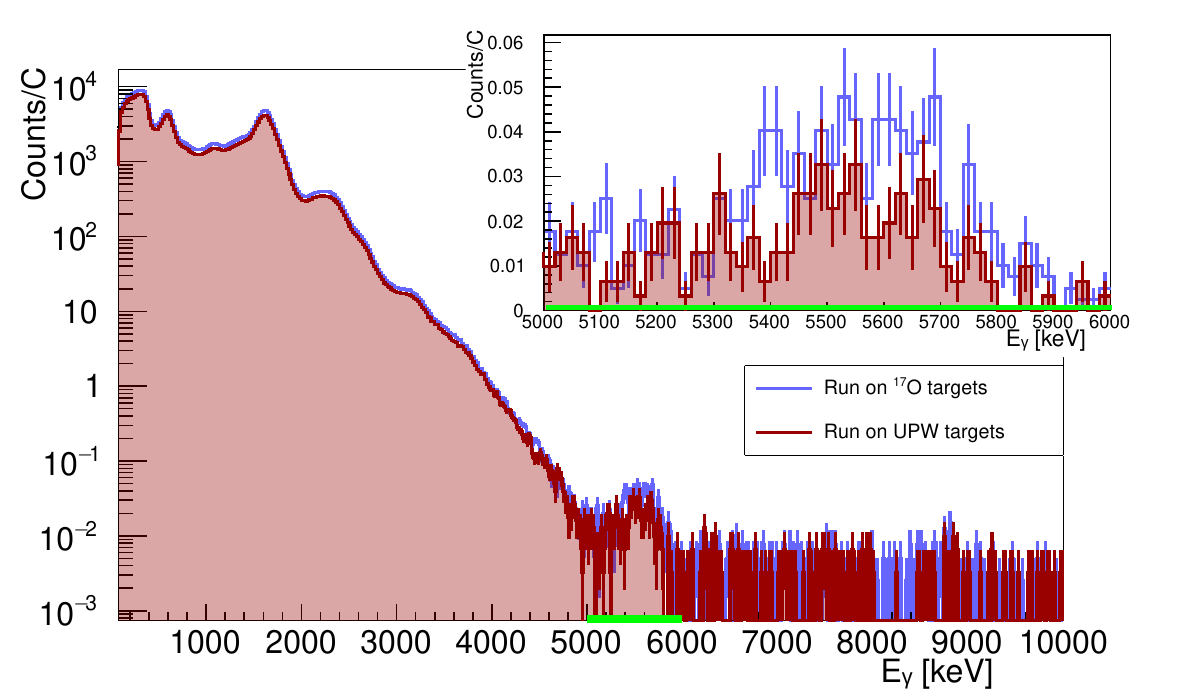}
   \caption{Comparison between the add-back spectra acquired on \isotope[17]{O} and UPW targets, in logarithmic scale. 
            A peak is visible in the energy region of interest, highlighted in green and shown in the insert in linear scale.}
   \label{fig:O18vO17}
 \end{figure}

\begin{figure}
   \centering
   \includegraphics[width=\columnwidth]{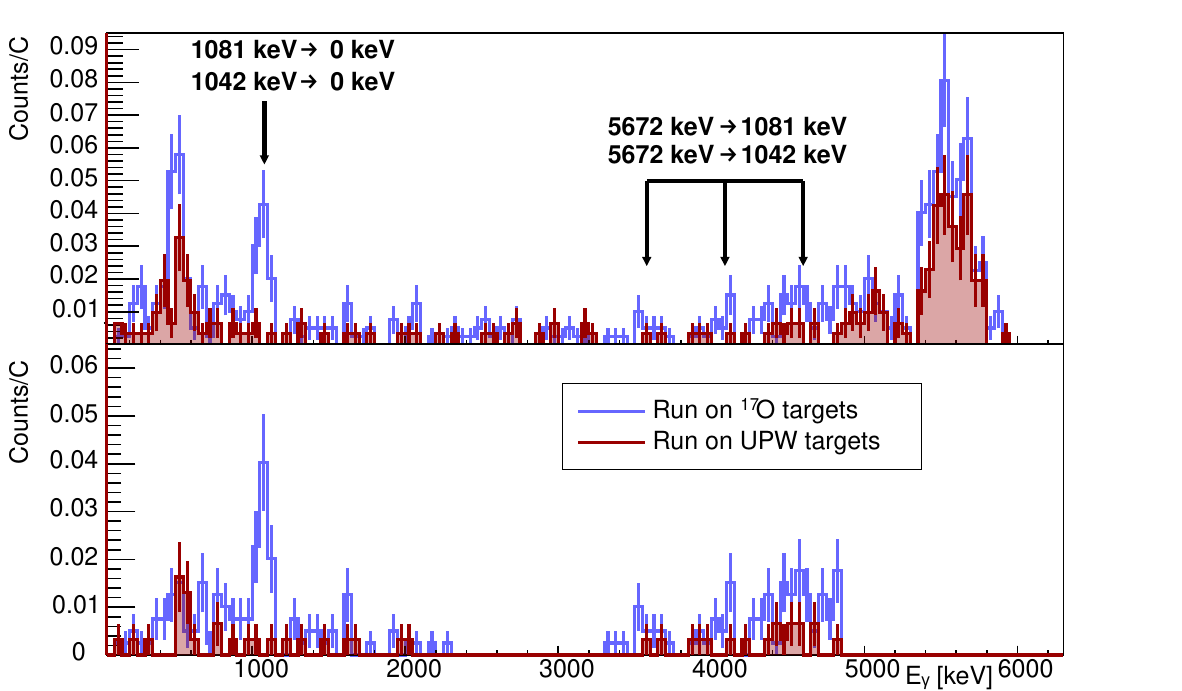}
   \caption{Coincidence method applied to the \isotope[17]{O} (blue) and UPW spectra (red). Top panel: events that add up to $5672$ keV; the energies corresponding to the primary and secondary \(\gamma\) of the transition to $1081$ keV are shown using arrows. Bottom panel: result of the secondary filter, only events belonging to the transition $5672$ keV $\to 1081$ keV are left in the chosen ROI; random coincidences survived the second filter and they were subtracted as estimated from the UPW measurement.}
   \label{fig:gate}
 \end{figure}

To monitor possible beam induced background in the ROI, targets were produced by performing
the anodic oxidation in a solution of ultra pure water (UPW) and water enriched in 
$^{18}$O at the 80\% level. The UPW targets had the same thicknesses as 
\isotope[17]{O} targets but negligible amount of $^{17}$O isotope. A comparison of the summed add-back spectra acquired on \isotope[17]{O} and UPW targets is shown in \autoref{fig:O18vO17}.  
A peak centered at the energy of interest, $E_{\gamma}\,=\,5672\,$keV,
is also visible in the UPW target spectra. 
This was attributed to a \isotope[2]{H} contamination
in the tantalum backings \cite{Asakawa-2020JVSTB}.
Due to the low BGO resolution, the \isotope[2]{H}(p,$\gamma$)\isotope[3]{He} reaction ($Q$-value = $5493.47508(6)$ keV \cite{Wang-2021}) 
peak cannot be distinguished from the $^{17}$O(p,$\gamma$)$^{18}$F resonance peak at present beam energy.
The deuterium contamination in tantalum was estimated to be of few ppm, too low to be eliminated via mechanical or chemical methods. The identification of beam induced background takes advantage of both the segmentation of the BGO detector and the 
knowledge of the decay scheme of the resonant state of interest 
of \isotope[18]{F} isotope \cite{Tilley_1995}.
This analysis is described in detail in \cite{Skowronski-2023JPhG},
and used in \cite{ferraro_high-efficiency_2018, piatti_first_2022, Zhang_19F_pg_2022}.
In short, we implemented the method as follows: 
first the events with total energy in the sum peak ROI were selected;
second, among these, we selected the events having deposited, in a single crystal, an energy corresponding to the 1081 and 1042 keV states
($3300\le\,E_{\gamma}\,\le4850$~keV), see \autoref{fig:gate}.
We applied the method to the simulations to obtain the gate efficiency, $24.3\%\pm0.7\%$.
This approach allows for complete discrimination between events belonging to the \isotope[17]{O}(p,$\gamma$)\isotope[18]{F}
reaction and those produced by the \isotope[2]{H}(p,$\gamma$)\isotope[3]{He} direct capture,
since the latter proceeds solely to the ground state emitting one $\gamma$ ray.
A downside of the method is a loss of statistics, since the states we gated on have an
overall intensity of 60.1(37)\%\cite{Tilley_1995}.
The background that survived the second gate is due to random coincidences, mimicking the cascade of interest.
We estimated and subtracted this contribution
applying the same gate analysis on runs acquired on UPW targets, as shown in \autoref{fig:gate}. 
 
To obtain the net yield of the $E_r = 64.5$ keV resonance we evaluated and subtracted the direct capture contribution to the observed count rate. 
An R-Matrix fit \cite{Azuma10-PRC} of all the available data \cite{Fox2005, Chafa2005, Chafa2007, Hager2012, Kontos-2012, DiLeva-2014, Buckner-2015} was performed to extrapolate the direct capture contribution to the astrophysical S-factor \footnote{$ S(E) = E \sigma(E) \exp[2\pi\eta(E)]$,
where $E$ is the energy in the center of mass frame, $\sigma(E)$ is the cross section and $\eta(E)$ is the Sommerfeld parameter~\cite{Iliadis07-Book}} down to the resonance energy \cite{Rapagnani2024}. The branching ratios for the capture to different excited states were found to be constant over the energy range $167 \le E \le 370 $ keV \cite{DiLeva-2014}. Therefore, those same branchings were included in the simulation code to get the efficiency for the gate analysis, applied to the direct capture case. By combining the direct capture S-factor from the R-matrix fit and the efficiency from the simulation, we infer a 0.04 reactions/C direct capture yield a $E_{\mathrm{p}} = 80$ keV. This contribution is subtracted from the measured yield at that beam energy to obtain the yield due to the resonant reaction. The direct capture contributes to 8\% of the measured experimental yield, see \autoref{eq:yield}. 
This result was verified performing one measurement at $E_{\mathrm{p}} = 265$ keV, that 
falls within literature data \cite{DiLeva-2014, Buckner-2015}, 
and one measurement at $E_{\mathrm{p}} = 142$ keV, to be compared with the R-Matrix output. 
The measured S-factors are: $S = (6.65\pm0.13)$ keVb and $S = (6.40\pm0.40)$ keVb, respectively.
These results are in agreement with previous experimental data and the new R-Matrix calculation.

The present result for the experimental yield is $Y_{\mathrm{exp}} = (0.50 \pm 0.10_{\mathrm{stat}}\pm 0.04_{\mathrm{syst}})$ reactions/C.
In the statistical uncertainty we included the contribution from the beam induced background, the direct capture subtraction and the composition of \isotope[17]{O} targets. 
The total systematic uncertainty accounts for uncertainty due to efficiency (3\%), branchings (6\% see \autoref{fig:figure1}), stopping power uncertainty (4\%) and charge integration (2\%). 
We also included the uncertainty of the resonance strength at $E_r = 183$ keV \cite{DiLeva-2014}, used as the reference to determine the \isotope[17]{O} isotopic abundance in targets.

Using the aforementioned yield we obtained a resonance strength of
$\omega\gamma_{(p,\gamma)}$ = (34 $\pm$ 7\textsubscript{stat} $\pm$ 3\textsubscript{syst})~peV. A comparison with literature data is presented in \autoref{tab:soa}.
The electron screening correction $f$ = 1.15 was derived considering the adiabatic approximation \cite{Assenbaum-1987} resulting in a $\omega\gamma_{(p,\gamma)}$\textsuperscript{bare} = (30 $\pm$ 6\textsubscript{stat} $\pm$ 2\textsubscript{syst})~peV.
However, a recent work suggests that the screening correction in case of narrow resonances is negligible \cite{Iliadis2023PhRvC}.
It must also be noted that a recent work reported stopping powers for proton in Ta higher by 12\% with respect to the SRIM database \cite{Moro-PRA20}. This would lead to an increase of the present effective stopping power, and consequently of the resonance strength, by 6\%.

The present result for the resonance strength is the first obtained by a direct
measurement and it is higher by a factor of $\approx$2 
than values reported in literature. 
Using $\Gamma_\gamma\,=\,$0.45(2) eV (the weighted mean of results in \cite{Parker-1968, Rolfs-1973, Berka-1977, Becker-1982}), $\Gamma_\alpha\,=\,$130(5) eV \cite{Mak-1980}, see \autoref{tab:soa}, and the present resonance 
strength, a $\Gamma$\textsubscript{p} = (39 $\pm$ 8\textsubscript{stat} $\pm$ 3\textsubscript{syst})~neV (corresponding to $\Gamma$\textsubscript{p}\textsuperscript{bare} = (34 $\pm$ 7\textsubscript{stat} $\pm$ 3\textsubscript{syst})~neV) was calculated, in excellent agreement with the previous LUNA result reported in \cite{Bruno-2016PhRvL}.
The weighted average of these two independent results yields $\Gamma_{\mathrm{p}}\,=\,35(5)$ neV, 
which is inconsistent with the previous literature values, i.e., 19(3) \cite{Iliadis_2010} and 14(2) \cite{Sergi-2015}.
The present work confirms the evaluation of the strength of the $64.5$ keV resonance in the \(\alpha\) channel
reported in \cite{Bruno-2016PhRvL} and the astrophysical consequences discussed in 
\cite{Lugaro-2017NatAs, Straniero-2017A&A}.

In summary, we reported the first direct measurements of the 64.5 keV 
resonance strength and the direct capture contribution at 142 keV in $^{17}$O(p,$\gamma$)$^{18}$F
reaction. To our knowledge this is the lowest strength value ever measured directly, corresponding to a cross section of 9.8 pb.
The deep underground location of LUNA, the improvements to the 
setup, and the application of the gate analysis allowed us to achieve an outstanding sensitivity, opening the path to future challenging measurements.
Our result for the resonance strength is roughly a factor of 2 higher than
previous values reported in literature, 
suggesting also a higher $\Gamma$\textsubscript{p}. The proton width calculated here is in excellent agreement with previous 
results by LUNA \cite{Bruno-2016PhRvL}, improving our understanding of the $^{16}$O/$^{17}$O ratio measured in red-giant stars \cite{Lebzelter-2015, Lebeltzer-2019, Hinkle-2016, DeNutte2017} and in O-reach pre-solar grains (for references see fig. 3 of \cite{Floss2016}). 
The full impact of the present new measurement on the uncertainty
of the rates of the \isotope[17]{O}(p,$\alpha$)\isotope[14]{N} and \isotope[17]{O}(p,$\gamma$)\isotope[18]{F} reactions
and their astrophysical implications will be discussed in a 
forthcoming work \cite{Rapagnani2024}.

\paragraph*{Acknowledgments} D. Ciccotti and the technical staff of the LNGS are
gratefully acknowledged for their help during setup construction and data taking. Dr. Sara Carturan from LNL and the chemistry laboratory staff of LNGS are acknowledged for the help with target preparation.
Financial support by INFN, the
Italian Ministry of Education, University and Research (MIUR) (PRIN2022 CaBS, CUP:E53D230023 and SOCIAL, CUP:I53D23000840006) and through the "Dipartimenti di eccellenza" project "Science of the Universe", the European Union (ERC Consolidator Grant project {\em STARKEY}, no. 615604, ERC-StG SHADES, no. 852016), (ELDAR  UKRI ERC StG (EP/X019381/1)) and (ChETEC-INFRA, no. 101008324), Deutsche For\-schungs\-gemeinschaft (DFG, BE 4100-4/1), the
Helm\-holtz Association (ERC-RA-
0016), the Hungarian National Research, Development and
Innovation Office (NKFIH K134197), the European Collaboration for Science and Technology (COST Action ChETEC, CA16117) and the Hungarian
Academy of Sciences via the Lend\"ulet Program LP2023-10. C. G. B., T. C., T. D. and M. A. acknowledge funding by STFC UK (grant no. ST/L005824/1).

For the purpose of open access, the authors have applied a Creative Commons Attribution (CC BY) license to any Author Accepted Manuscript version arising from this submission.

\end{document}